\documentclass[12pt]{iopart}

\usepackage{graphicx}
\usepackage{color}

\definecolor{tcr}{rgb}{0, 0.0, 0.0}

\newcommand{\tcr}[1]{\textcolor{black}{#1}}

\begin{document}

\title[]{Quantum non-Gaussianity criteria based on vacuum probabilities of original and attenuated state}

\author{Jarom\'{\i}r Fiur\'{a}\v{s}ek, Lukáš Lachman, Radim Filip}
\address{Department of Optics, Palack\'{y} University Olomouc, 17. listopadu 1192/12, 771 46 Olomouc, Czech Republic}

\ead{fiurasek@optics.upol.cz, filip@optics.upol.cz}

\vspace{10pt}

\begin{abstract}
Quantum non-Gaussian states represent an important class of highly non-classical states whose preparation requires quantum operations or measurements beyond the class of Gaussian operations \tcr{and statistical mixing}. 
Here we derive criteria for certification  of quantum non-Gaussianity based on probability of vacuum in the original quantum state and a state transmitted through a lossy channel with transmittance $T$. 
We prove that the criteria hold for arbitrary multimode states, which is important for their applicability in experiments with broadband sources and single-photon detectors.
Interestingly, our approach allows to detect quantum non-Gaussianity  using only one photodetector instead of complex multiplexed photon detection schemes, at the cost of increased experimental time.  
 We also formulate a quantum non-Gaussianity  criterion based on the vacuum probability  and mean photon number of the state and we show that this criterion is closely related to the criteria based on pair of vacuum probabilities.
We illustrate the performance of the obtained criteria on the example of realistic imperfect single-photon states modeled as a mixture of vacuum and single-photon states with background Poissonian noise. 
\end{abstract}

\section{Introduction}

 Characterization and classification of quantum states of light represents one of the main topics in quantum optics since its early days and has remained an important subject of research today. 
 The non-classical states of light are traditionally defined as states that cannot be expressed as classical mixtures of coherent states. For nonclassical states, the 
 Glauber-Sudarshan $P$ representation \cite{Glauber1963,Sudarshan1963} loses the properties
 of ordinary probability distribution and becomes a generalized quasidistribution. Photodetection events from such states cannot be explained by classical coherence theory \cite{Mandel1986} 
 and, therefore, they have potential to overcome classical optical technology. By considering all
$s$-parameterized quasidistributions \cite{Perina1984} from the $P$-representation function to Husimi $Q$-function one can define a degree of nonclassicality in terms of the largest value of $s$ 
for which the $s$-parameterized quasidistribution remains a classical probability distribution \cite{Lee1991}. A more stringent criterion of non-classicality is based on the negativity of Wigner function.
 In current quantum optics experiments and continuous variable quantum information processing, states with Gaussian Wigner function that encompass quadrature squeezed states 
 in addition to coherent \tcr{and thermal} states can be relatively easily generated, processed and detected. Although the Gaussian states and Gaussian operations represent a very valuable resource 
 for quantum sensing and quantum communication \cite{Weedbrook2012},  they are not sufficient for several key tasks in optical quantum technologies, such as entanglement distillation 
 \cite{Giedke2002,Eisert2002,Fiurasek2002,Takahashi2010,Kurochkin2014,Ulanov2015}, 
quantum error correction \cite{Niset2009,Hu2019,Ma2020}  or quantum computing \cite{Bartlett2002}.

 In order to identify and characterize highly-nonclassical quantum states beyond the Gaussian states, the class of quantum non-Gaussian states of light has been introduced \cite{Filip2011}. 
 A quantum state \tcr{$\hat{\rho}$} is said to be quantum non-Gaussian if it is not a Gaussian state or a statistical mixture of Gaussian states. The concept of quantum non-Gaussianity \tcr{is thus similar to} 
 the concept of nonclassicality \tcr{because in both cases we define some convex sets of states and are interested in states that do not belong to those convex sets.} 
It should be noted that the class of quantum non-Gaussian states is strictly larger than the class of states with negative Wigner function \cite{Filip2011}. 
 For example, all mixtures of vacuum and single-photon states $p|1\rangle \langle 1|+(1-p)|0\rangle \langle 0|$ with $p>0$ are quantum non-Gaussian, but the state exhibits 
 negative Wigner function only if $p>1/2$. 
 
Various criteria and witnesses to detect and certify quantum non-Gaussian states have been proposed in the literature based on photon number distribution
 \cite{Filip2011,Lachman2013,Kuhn2018}, quadrature operators and homodyne detection \cite{Park2017, Happ2018}, and properties of phase space distributions \cite{Genoni2013,Hughes2014}. 
 The quantum non-Gaussian character of approximate single-photon states conditionally prepared by detection of an idler photon from a correlated photon pair generated in the process of spontaneous parametric downconversion, 
 or emitted by quantum dots or trapped ions has been studied experimentally \cite{Jezek2011,Straka2014,Higginbottom2016}. 
 A quantitative measure of quantum non-Gaussianity has been proposed based on the general concepts of resource theory where the Gaussian states are treated as free states \cite{Takagi2018,Albarelli2018,Park2019}.
 The concept of quantum non-Gaussiannity has been recently extended to genuine $n$-photon quantum non-Gaussianity \cite{Lachman2019}, where one considers squeezed 
 and displaced superpositions of Fock states up to Fock state $|n-1\rangle$ as free states,  instead of just the ordinary Gaussian states. 
The connection between quantum non-Gaussianity and secure quantum communication has been studied \cite{Lee2019} 
 and quantum non-Gaussianity of photon subtracted squeezed states has been analyzed \cite{Jezek2012,Song2013}. 
 Also quantum non-Gaussianity of photons in optical and optomechanical conversions has been investigated \cite{Baune2014, Rakhubovsky2017}.

 The quantum non-Gaussianity of an optical quantum state can be tested with a simple Hanbury-Brown Twiss experimental setup that is commonly used for  measurement of the $g^{(2)}$ factor \cite{Grangier1986,Bocquillon2009}. 
 In this setup, the optical signal is split 
 on a balanced beam splitter and each output port is measured with a single-photon detector that distinguishes the presence and absence of photons \cite{Lachman2013,Jezek2011}. 
 The quantum non-Gaussianity criterion can then be expressed in terms of the directly measured probability of coincidence clicks and single detector clicks. Interestingly, one can equivalently formulate this criterion in terms 
 of the probability of vacuum in the incident optical quantum state and a probability of vacuum in a quantum state transmitted through a purely lossy channel with transmittance $T=1/2$ \cite{Lachman2013}. 
 However, we do not have to restrict ourselves to $50\%$ losses and we can consider a channel with any transmittance $T$. 
 This approach based on no-click probablities has been previously used to  derive and experimentally verify nonclassicality criteria 
suitatable for large number of emitters \cite{Filip2013, Lachman2016,Moreva2017,Obsil2018,Qi2018}.

In the present paper, we extend this approach to certification of quantum non-Gaussianity 
and we derive a class of quantum non-Gaussianity criteria based on knowledge of probability of vacuum in the original state and in a state transmitted through a lossy channel with an arbitrary transmittance $T$. 
 Our results generalize the quantum non-Gaussianity criteria  obtained for $T=1/2$ in Ref. \cite{Lachman2013}. We provide a detailed mathematical proof that the resulting criteria are fully applicable to general multimode states 
 with arbitrary number of modes. This is crucial for their practical utilization, because in most photon counting experiments it is impossible to ensure that only a single  mode is detected 
\tcr{unless one employes a sophisticated and technically demanding temporal filtering via quantum pulse gating \cite{Eckstein2011,Reddy2018}. }
 We show that the class of quantum non-Gaussianity criteria obtained in the present work 
 is strictly stronger than the original criterion for $T=1/2$, because there exist quantum non-Gaussian states that cannot be detected by the criterion for $T=1/2$ but can be detected by the present generalized 
 criteria, when choosing a suitable $T\neq 1/2$. Furthermore, we derive a quantum non-Gaussianity criterion based on the knowledge of vacuum probability and mean photon number of the state and we show that 
 this latter criterion can be interpreted as  a specific case of the above class of criteria considered in the limit $T \rightarrow 0$. We illustrate application of the obtained criteria on the example of  attenuated single-photon state with Poissonian background noise, 
which models a realistic imperfect single-photon state. In the concluding part of the paper we discuss the scaling of the number of measurement events required for reliable confirmation of quantum non-Gaussianity and compare  performance of two setups that involve either one or two single-photon detectors.

\section{Quantum non-Gaussianity criterion based on vacuum probabilities}

In this section we derive a criterion of quantum non-Gaussianity for a state $\hat{\rho}$ based on the probability of vacuum $p_0=\langle 0|\hat{\rho}|0\rangle$ and probability $q_0(T)$ of vacuum after transmission through 
a lossy quantum channel $\mathcal{L}$ with transmittance $T$, $q_0(T)=\langle 0|\mathcal{L}_T(\hat{\rho})|0\rangle$. Note that
\begin{equation}
q_0(T)=\sum_{n=0}^{\infty} p_n (1-T)^n,
\end{equation}
where $p_n=\langle n|\hat{\rho}|n\rangle$ is the photon number distribution of state $\hat{\rho}$. The quantum non-Gaussianity criterion has the form of an upper bound on $q_0(T)$ 
that is achievable by Gaussian states and their mixtures for a given fixed $p_0$. If the experimentally detected $q_0(T)$ exceeds this bound, then the state is certified as quantum non-Gaussian. 
Equivalently, the criterion can be formulated in terms of the quantum non-Gaussianity witness
\begin{equation}
W=q_0(T)-\lambda p_0
\label{Wdefinition}
\end{equation}
 for a suitable Lagrange multiplier $\lambda$. 
An upper bound on the witness $W$ achievable by Gaussian states and their mixtures can be established, and if this bound is exceeded, the quantum non-Gaussianity of the state is confirmed. 

\begin{figure}[t]
\centerline{\includegraphics[width=0.85\linewidth]{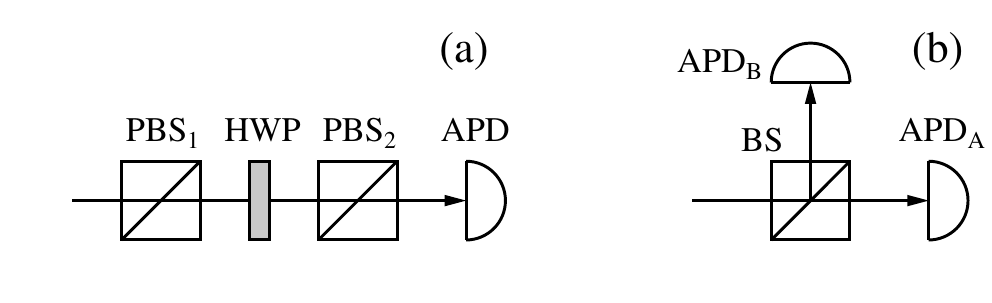}}
\caption{Optical schemes for characterization of quantum non-Gaussianity. 
(a) Single-APD scheme: the input signal passes through a polarizing beam splitter PBS$_1$ to ensure that it is fully polarized and is then attenuated by combination of a rotated half-wave plate HWP and another PBS. 
This configuration realizes a tunable  lossy channel \tcr{${\mathcal{L}}_T(\hat{\rho})$} with transmittance $T$ controlled by the rotation angle of HWP.
The output signal is detected with avalanche photodiode APD. (b) Double-APD scheme: the input signal is split on a beam splitter with transmittance $T$ and detected by two avalanche photodiodes APD. }
\label{fig0}
\end{figure}

In quantum optics setting, the probabilities $p_0$ and $q_0(T)$ can be \tcr{sequentially} measured with linear optics and a \emph{single} avalanche photodiode APD that distinguishes between the presence or absence of photons. 
A schematic experimental setup is illustrated in Fig. 1(a), where a horizontally polarized light beam propagates through a variable attenuator formed by a combination of a half-wave plate and a polarizing beam splitter,
followed by the single-photon detector APD. Any transmittance $T$ can be set by rotation of the half-wave plate.
\tcr{The probability $p_0$ is then given by the probability of no-clicks of the APD when we make the channel fully transparent and set $T=1$. In order to determine $q_0(T)$ we have to set the chosen transmittance $T$ and again measure the probability of no-clicks of the APD.}
 An alternative setup with two APDs  is shown in Fig. 1(b). 
Here the input signal beam is split on a beam splitter with transmittance $T$ and APDs are placed on both output ports of the beam splitter. The probability $p_0$ corresponds to the probability that none of the APDs clicks, 
while the probability $q_0(T)$ is given by the probability that APD$_A$ does not click, irrespective of the measurement outcome of APD$_B$. 
Note that in this setting we can simultaneously measure also $q_0(1-T)$ by recording the probability of no-clicks of APD$_{B}$.

The avalanche photodiodes exhibit limited detection probability $\eta$. Such detectors can be modeled as a sequence of a lossy channel with transmittance $\eta$ and an ideal on-off detector 
that perfectly distinguishes the presence and absence of photons.
In the scheme on Fig 1(a), limited detection efficiency implies that instead of quantum non-Gaussianity of state $\hat{\rho}$ one probes the quantum non-Gaussianity of state $\mathcal{L}_\eta(\hat{\rho})$. 
Since lossy channels map Gaussian states and their mixtures onto Gaussian states and their mixtures, proving quantum non-Gaussianity of $\mathcal{L}_\eta(\hat{\rho})$ implies that 
also the original state $\hat{\rho}$ is quantum non-Gaussian. The quantum non-Gaussianity criteria derived in the present paper can therefore be directly 
applied to measurements with setup in Fig. 1(a) without any need for calibration of the detection efficiency. 
This changes for the setup in Fig. 1(b), where the two APDs can exhibit different detection efficiencies $\eta_A$ and $\eta_B$. 
The fractions of the signal that are effectively detected by the two APDs can be expressed as $\eta_A T$ and $\eta_B (1-T)$,
yielding an effective splitting ratio 
\begin{equation}
\tilde{T}=\frac{\eta_A T}{\eta_A T+\eta_B(1-T)}.
\end{equation}
Since the quantum non-Gaussianity criteria studied in this work depend on $T$, utilization of setup in Fig. 1(b) would require calibration of detection efficiencies and determination of their relative ratio $\eta_A/\eta_B$, or, equivalently,
determination of the relative ratio of signals detected by each APD, $\frac{\eta_A T}{\eta_B(1-T)}$.

A quantum non-Gaussianity criterion based on $p_0$ and $q_0(T)$ can be derived  by calculating the maximum of the witness (\ref{Wdefinition}) that is attained by Gaussian states and their mixtures. 
The witness can be expressed as $W=\mathrm{Tr}[\hat{\rho} \hat{W}_T]$, where $\hat{W}_T=(1-\lambda)|0\rangle\langle 0|+\sum_{n=1}^{\infty}(1-T)^n |n\rangle\langle n|$. 
It follows that the maximum is achieved by a pure Gaussian state, because for any mixture $\hat{\rho}=\sum q_k |\psi_k\rangle\langle \psi_k|$ 
we have\tcr{ $\mathrm{Tr}[\hat{\rho} \hat{W}_T] \leq \max_k \langle\psi_k|\hat{W}_T|\psi_k\rangle$}. Let $\hat{r}=(\hat{x},\hat{p})$ denote the vector of quadrature operators. Any single-mode Gaussian state is fully characterized by
the coherent displacement vector $\vec{d}=(d_x,d_p)=\langle \hat{r}\rangle$ and by a covariance matrix $\gamma_{jk}=\langle \Delta \hat{r}_j \Delta \hat{r}_k+ \Delta \hat{r}_k \Delta \hat{r}_j\rangle $.
Since phase shifts do not change photon number distribution, we can assume without loss of generality that the covariance matrix is diagonal. For pure Gaussian states $\det\gamma=1$ and we can write $\gamma=\mathrm{diag}(V,1/V)$,
where $V$ is a normalized quadrature variance that satisfies $V=1$ for vacuum and coherent states. 
The probability of vacuum $p_0$ can be conveniently calculated from the Husimi $Q$-function, which is defined as $Q(\alpha)=\frac{1}{\pi}\langle \alpha|\hat{\rho}|\alpha\rangle$, where $|\alpha\rangle$ denotes a coherent state. 
For Gaussian states, the $Q$-function is a Gaussian probability distribution with mean $\vec{d}/\sqrt{2}$ and covariance matrix $\frac{1}{2}(\gamma+I)$, where $I$ is the identity matrix. We have $p_0=\pi Q(0)$, 
which for the considered pure Gaussian state explicitly yields 
\begin{equation}
p_{0}=\frac{2\sqrt{V}}{V+1}\exp\left(-\frac{d_x^2}{V+1}-\frac{V d_p^2}{V+1}\right).
\label{p0formula}
\end{equation}
After passing through a lossy channel with transmittance $T$ the displacement and covariance matrix change as follows,
\begin{equation}
\vec{d}\rightarrow \sqrt{T}\vec{d}, \qquad  \gamma\rightarrow T\gamma+(1-T)I.
\end{equation}
With these formulas at hand its is straightforward to calculate the  probability of vacuum $q_0(T)$ for a state transmitted through the lossy channel,
\begin{equation}
\fl
q_0(T)=\frac{2\sqrt{V}}{\sqrt{(TV+2-T)(T+2V-TV)}}\exp\left(-\frac{T d_x^2}{TV+2-T}-\frac{TV d_p^2}{T+2V-TV}\right).
\label{q0Tformula}
\end{equation}

We now determine the optimal Gaussian state that maximizes the quantum non-Gaussianity witness $W=q_0(T)-\lambda p_0$. The extremality conditions read
\begin{equation}
\fl
\frac{\partial q_0(T)}{\partial d_x}-\lambda \frac{\partial p_0}{\partial d_x}=0, \qquad 
\frac{\partial q_0(T)}{\partial d_p}-\lambda \frac{\partial p_0}{\partial d_p}=0, \qquad 
\frac{\partial q_0(T)}{\partial V}-\lambda \frac{\partial p_0}{\partial V}=0. 
\label{dW}
\end{equation}
By eliminating the Lagrange multiplier $\lambda$ from the first two formulas in Eq. (\ref{dW}) we obtain the condition
\begin{equation}
\frac{\partial q_0(T)}{\partial d_x}\, \frac{\partial p_0}{\partial d_p}-\frac{\partial q_{0}(T)}{\partial d_p} \, \frac{\partial p_0}{\partial d_x}=0,
\end{equation}
which yields
\begin{equation}
\frac{8d_x d_p T (1-T)V(1-V)}{(T+2V-TV)(TV+2-T)} p_0 q_0(T)=0.
\end{equation}
This can be satisfied only if either $d_x=0$ or $d_p=0$, or if $V=1$. This latter condition corresponds to input coherent state and for these states we can set $d_p=0$ without loss of generality. 
Therefore, it suffices to consider a two-parametric class of pure squeezed coherent states with $d_p=0$. The second formula in Eq. (\ref{dW}) is then satisfied automatically
and the elimination of $\lambda$ from the first and third formulas in Eq. (\ref{dW}) leads to
\begin{equation}
\frac{\partial q_0(T)}{\partial d_x}\, \frac{\partial p_0}{\partial V}-\frac{\partial q_{0}(T)}{\partial V} \, \frac{\partial p_0}{\partial d_x}=0.
\label{aVcondition}
\end{equation}
On inserting the explicit expressions for $q_0(T)$ and $p_0$ into Eq. (\ref{aVcondition}), we obtain after some algebra
\begin{equation}
\fl
\frac{2d_x(1-T)T[2-(2+4d_x^2)V^2-T(1-V)(1+2d_x^2V+V^2)]}{V(1+V)^2(2+TV-T)^2(T+2V-TV)}p_0 q_0(T)=0.
\end{equation}
This latter condition can be satisfied either if $d_x=0$ or if the term in the square brackets in the numerator is equal to $0$. The case
$d_x=0$ corresponds to squeezed vacuum state with super-Poissonian statistics. One can expect that this class of states is not optimal for the considered criterion, and
this is rigorously proved in the Appendix A. If follows that the term in the square brackets must vanish, which yields a formula for the optimal $d_x^2$,
\begin{equation}
d_{x,\mathrm{opt}}^2=\frac{1-V^2}{2V}\frac{2-T+TV}{2V-TV+T}.
\label{dxopt}
\end{equation}
Note that $V\leq 1$ must hold, i.e. the state must be an amplitude squeezed coherent state, otherwise a positive solution for $d_x^2$ does not exist. 
On inserting the expression (\ref{dxopt}) for the optimal $d_x^2$ back to the formulas (\ref{p0formula}) and (\ref{q0Tformula}) for $p_0$ and $q_{0}(T)$, where we set \tcr{$d_p=0$}, 
we obtain parametric expression for the maximum $q_0(T)$ achievable by Gaussian states and their mixtures for a given $p_0$ and $T$,
\begin{eqnarray}
\fl
p_0(V,T)=\frac{2\sqrt{V}}{V+1}\exp\left[-\frac{(1-V)(2-T+TV)}{2V(2V-TV+T)}\right], \nonumber \\
\fl
q_0(V,T)=\frac{2\sqrt{V}}{\sqrt{(TV+2-T)(T+2V-TV)}}\exp\left[-\frac{T(1-V^2)}{2V(2V-TV+T)}\right],
\label{qpsinglemodeoptimal}
\end{eqnarray}
where $V\in(0,1]$.

\begin{figure}[t]
\centerline{\includegraphics[width=0.95\linewidth]{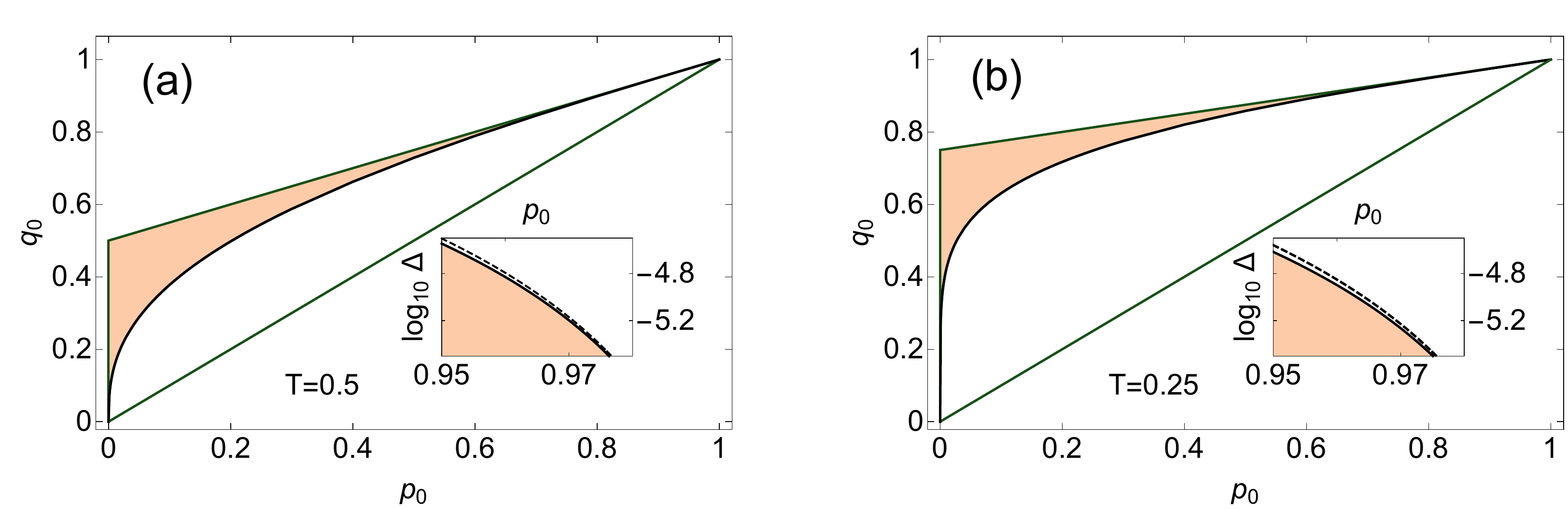}}
\caption{Thresholds for detection of quantum non-Gaussian states. 
The black lines indicate the dependence of the threshold probability $q_0(V,T)$ on $p_0(V,T)$ for $T=0.5$ (a) and $T=0.25$ (b).
A quantum state is certified to be quantum non-Gaussian if its probability pair $p_0$, $q_0$ corresponds to a point in the orange area. 
The green lines present the physical boundary in the space of the probabilities $p_0$ and $q_0$. 
The insets in both figures focus on the region of strongly attenuated states and show the threshold for  deviation from the physical boundary   $\Delta=1-T(1-p_0)-q_0 $. 
Note that $\Delta=0$ for  boundary states formed by mixtures of vacuum and single-photon states. 
Whereas the solid lines in the insets depict the exact thresholds, the dashed lines represent the approximate threshold condition specified in Eq. (\ref{approx}). 
}
\end{figure}

The formulas (\ref{qpsinglemodeoptimal}) represent the main result of the present paper and they generalize the results for $T=\frac{1}{2}$ derived in Ref. \cite{Lachman2013}. 
We note that the  nonclassicality criteria derived previously for the same setup have much simpler form $q_0(T) > p_0^T$ \cite{Filip2013,Obsil2018}.
In Fig. 2 we plot the dependence of $q_0(V,T)$ on $p_0(V,T)$ for $T=0.5$ and $T=0.25$.
The area of pairs $(p_0,q_0(T))$ which certify the quantum non-Gaussianity is indicated by orange color. The range of physically allowed $q_0(T)$ is lower bounded by $q_0(T)=p_0$ 
and upper bounded by $q_0(T)=1-T(1-p_0)$.  The upper bound follows from the inequality 
\[
\sum_{n=1}^\infty(1-T)^n p_n\leq \sum_{n=1}^\infty (1-T)p_n=(1-T)(1-p_0), 
\]
and is saturated by mixtures of vacuum and single-photon states,  $\hat{\rho}=p_0|0\rangle\langle 0|+(1-p_0)|1\rangle\langle 1|$. 
All such mixtures are therefore detectable to be the quantum non-Gaussian states.  
If we experimentally obtain values $p_{0,\mathrm{exp}}$ and $q_{0,\mathrm{exp}}(T)$, then we can apply the criterion by finding $V$ such that 
$p_{0,\mathrm{exp}}=p_0(V,T)$ and then we can compare  $q_0(V,T)$  with $q_{0,\mathrm{exp}}$. Note that  $p_0(V,T)$ is a monotonously increasing function of $V$, therefore the 
equation $p_{0,\mathrm{exp}}=p_0(V,T)$ has a unique solution $V$. However, we should take into account that both the experimentally determined $p_{0,\mathrm{exp}}$ 
and $q_{0,\mathrm{exp}}(T)$ will exhibit some statistical uncertainty. It is therefore more appropriate and advantageous to make use of the quantum non-Gaussianity witness (\ref{Wdefinition}).
 The Lagrange multiplier $\lambda$ can be  calculated  from the extremality conditions (\ref{dW}) and we obtain
\begin{equation}
\lambda(V,T)=\frac{(V+1)T}{TV+2-T} \frac{q_0(V,T)}{p_0(V,T)}.
\end{equation}
For this $\lambda$, the maximum of $W$ achievable with Gaussian states and their mixtures is given by 
\begin{equation}
W_{G}(V,T)=q_0(V,T)-\lambda(V,T)p_0(V,T)=\frac{2(1-T)}{TV+2-T}q_0(V,T).
\label{WGthreshold}
\end{equation}
The quantum non-Gaussianity of a state is certified if the experimentally determined witness $W_{\mathrm{exp}}=q_{0,\mathrm{exp}}(T)-\lambda(V,T)p_{0,\mathrm{exp}}$ exceeds the maximum $W_G(V,T)$. 
Note that for any $T$ we have a one-parametric class of witnesses, each specified by the value of $V$. When processing the experimental data, the parameter $V$ can be optimized 
to maximize the confidence of the experimental certification of the quantum non-Gaussianity. A more detailed discussion of this issue is provided in Section 5 below.
 
Importantly, the quantum non-Gaussianity criteria derived in this section are not restricted to single-mode states and are applicable to arbitrary multimode states. This is important for practical applicability of the criteria to 
 experiments with broadband single-photon sources and broadband single-photon detectors such as avalanche photodiodes.  
A detailed technical proof of the validity of the  criteria for multimode states is provided in Appendix B. Here we outline the main steps of the proof. We base the proof on the maximization of the witness $W$ 
over all multimode Gaussian states. First, we observe that it suffices to consider pure product multimode Gaussian states.  Then we focus on the two-mode case and we explicitly prove that the optimal two-mode Gaussian 
state that maximizes $W$ is a product of the optimal single-mode state and a vacuum state. Finally, we use mathematical induction to extend this result to arbitrary number of modes.

\section{Quantum non-Gaussianity criterion based on mean photon number}
In this section we derive a quantum non-Gaussianity criterion based on probability of vacuum $p_0$ and mean photon number $\bar{n}=\sum_{n=0}^\infty n p_n$. 
Our starting point is the following identity that connects the mean photon number $\bar{n}$ and the function $q_0(T)$,
\begin{equation}
\bar{n}=\sum_{n=0}^\infty n p_n = \left.- \frac{d q_0(T)}{dT}\right|_{T=0}= \lim_{T\rightarrow 0}\frac{1-q_0(T)}{T}. 
\end{equation}
This formula suggests that the criterion based on $p_0$ and $\bar{n}$ can be obtained as a limit case of the criteria based on $p_0$ and $q_0(T)$, when $T\rightarrow 0$. Specifically,
we show that the quantum state is quantum non-Gaussian if for a given $p_{0}$ the mean photon number $\bar{n}$ is smaller than certain threshold $\bar{n}_{\mathrm{th}}$. 
Let $\mathcal{G}_0$ denote the set of Gaussian states and their mixtures with probability of vacuum state equal to $p_{0}$. The threshold $\bar{n}_{\mathrm{th}}$ is then given by
\begin{equation}
\fl
\bar{n}_{\mathrm{th}}= \min_{\mathcal{G}_0}\lim_{T\rightarrow 0}\frac{1-q_0(T)}{T}=\lim_{T\rightarrow 0}\frac{1-\max_{\mathcal{G}_0} q_0(T)}{T}=\lim_{T\rightarrow 0}\frac{1- q_0(V(T),T)}{T},
\end{equation}
where the function $V(T)$ is implicitly defined by $p_{0}(V(T),T)=p_0$. It holds that
\begin{equation}
\bar{n}_{\mathrm{th}}= -\left.\left [\frac{\partial q_0(V,T)}{\partial T}+\frac{\partial q_0(V,T)}{\partial V} \frac{d V(T)}{d T} \right]\right|_{T=0}.
\end{equation}
Since $q_0(V,0)=1$ for any $V$, we have 
\begin{equation}
\left. \frac{\partial q_0(V,T)}{\partial V}  \right|_{T=0}=0.
\end{equation}
Consequently, the explicit dependence of $V(T)$ on $T$ is irrelevant and we obtain
\begin{equation}
\bar{n}_{\mathrm{th}}=\frac{(1-V)(1+2V-V^2)}{4V^2}.
\end{equation}
This, together with
\begin{equation}
p_{0}(V,0)=\frac{2\sqrt{V}}{V+1}\exp\left[-\frac{1-V}{2V^2}\right],
\end{equation}
defines the parametric dependence of the threshold $\bar{n}_{\mathrm{th}}$ on the vacuum probability $p_{0}$. If for a given $p_{0}(V,0)$ we observe $\bar{n}< \bar{n}_{\mathrm{th}}$ then 
the state is quantum non-Gaussian. The criterion based on the vacuum probability $p_0$ and the mean photon number $\bar{n}$ is suitable for measurements where both these quantities 
can be efficiently estimated from the collected data.  The (phase randomized) balanced homodyne detection or eight-port homodyne detection represent two examples of such measurement configurations \tcr{\cite{Munroe1995,Welsch1999}}.

\tcr{The above derived criteria based on $p_0$ and $q_0$ or $p_0$ and $\bar{n}$  can be generalized by considering an arbitrary Gaussian unitary transformation $\hat{U}_G$. 
The criteria can be equivalently applied to a transformed state $\hat{\rho}'=\hat{U}_G \hat{\rho} \hat{U}_G^\dagger$ and if this latter state is certified as quantum non-Gaussian then also the original state $\hat{\rho}$ is proved to be quantum non-Gaussian. 
This approach becomes particularly relevant if we characterize the state by phase sensitive homodyne detection, because in this case the Gaussian unitary squeezing and/or displacement operation can be included in the post-processing of the experimental homodyne data \cite{Jezek2012}.
If we consider states such as photon added or subtracted squeezed coherent states, then a suitable Gaussian unitary transformation $\hat{U}_G$ may remove the Gaussian envelope of the state and preserve only the non-Gaussian core \cite{Menzies2009} whose quantum non-Gaussianity may be more easily
certified by our criteria \cite{Jezek2012}.  }

\section{Attenuated single-photon state with Poissonian background noise}
\begin{figure}[t]
\centerline{\includegraphics[width=0.6\linewidth]{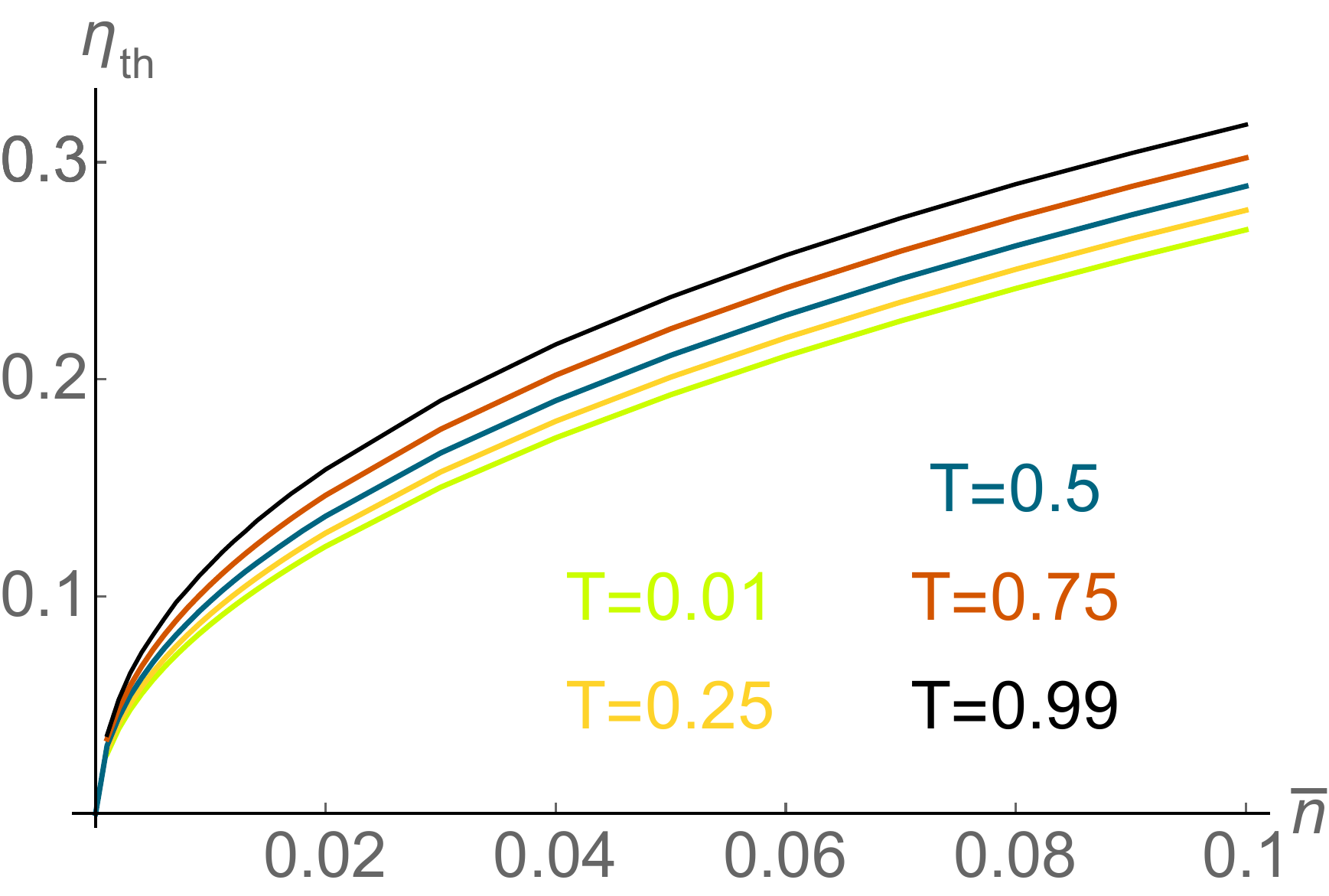}}
\caption{Certification of quantum non-Gaussianity of attenuated single-photon states with background noise. The threshold single-photon fraction $\eta_\mathrm{th}$ is plotted as a function of the mean number of noisy photons 
$\bar{n}$ for 5 different transmittances  $T$ in the detection scheme. Quantum non-Gaussianity of the state can be certified if $\eta$ exceeds the plotted threshold $\eta_{\mathrm{th}}$. }
\label{FigTdependenceModel}
\end{figure}

In this section we will illustrate the power of the above derived quantum non-Gaussianity criteria on the important example of approximate single-photon states.
Current single-photon sources can produce states with density matrix approaching $\hat{\rho}_{\eta}=\eta |1\rangle \langle 1 |+(1-\eta)|0\rangle \langle 0 |$, where $\eta$ is a product of the efficiency of the source, 
the collection efficiency and the quantum efficiency of the detector. 
In a more realistic scenario, background noise deteriorates the state \tcr{$\hat{\rho}_{\eta}$} and the resulting density matrix can be expressed as
\begin{equation}
   \tcr{ \hat{\rho}=\hat{\rho}_{\eta}\otimes \hat{\rho}_{\bar{n}},}
\label{rhodefinition}
\end{equation}
where we assume the experimentally relevant case with \tcr{$\hat{\rho}_{\bar{n}}$} obeying Poissonian photon-number distribution characterized by the mean number of noisy photons $\bar{n}$. 
Then, the involved probabilities read 
\begin{equation}
p_0=(1-\eta)\exp(-\bar{n}), \qquad q_0(T)=(1-\eta T)\exp(-\bar{n}T).
\label{p0q0attenuated} 
\end{equation}
The criteria  (\ref{qpsinglemodeoptimal}) can detect quantum non-Gaussianity of the state (\ref{rhodefinition}) provided that $\eta$ exceeds certain threshold value $\eta_{\mathrm{th}}$ that depends on $\bar{n}$ and $T$.
In Fig.~ \ref{FigTdependenceModel} we plot the dependence of this threshold value $\eta_{\mathrm{th}}$ on $\bar{n}$ for several values of $T$. This figure illustrates that the criteria for different $T$ 
are not equivalent and for the considered class of states (\ref{rhodefinition}) the criteria become most powerful for small $T$, where they can certify the quantum non-Gaussianity for a largest range of parameters $\eta$ and $\bar{n}$. 
We note that this behavior is not universal and one can also construct states whose quantum non-Gaussianity can be certified only using criteria with $T>1/2$. 
Importantly, we can conclude that the quantum non-Gaussianity criteria derived in the present work are more powerful then the specific criterion obtained for $T=1/2$.

The emerging single-photon sources often produce states that are significantly affected by losses. It is therefore interesting to investigate detection of quantum non-Gaussianity 
of highly attenuated states  (\ref{rhodefinition})  with $\eta\ll 1$ and $\bar{n}\ll 1$.  For such states both the probabilities $p_0$ and $q_0$ will be close to $1$. 
We can approximate the threshold (\ref{qpsinglemodeoptimal}) in this region of probabilities to understand the ability to certify quantum non-Gaussianity for this limit. 
By performing the Taylor expansion of $q_0(V,T)$ and $p_0(V,T)$ in Eq. (\ref{qpsinglemodeoptimal}) around the point $V=1$ we obtain
 \begin{eqnarray}
  1-q_0(V,T)\approx \frac{T}{2}(1-V), \nonumber \\
\frac{1-T}{T}-\frac{q_0(V,T)}{T}+p_{0}(V,T)\approx \frac{(1-T)(2-T)}{12}(1-V)^3.
\label{pqTaylor}
 \end{eqnarray}
 These simplified expressions allow us to exclude the parameter $V$ and establish an explicit approximate quantum non-Gaussianity criterion based on  $q_0$ and $p_0$. 
 After some algebra, we find that the state is quantum non-Gaussian if the following inequality holds,
 \begin{equation}
     (1-q_0)^3 > \frac{3T^2}{2(1-T)(2-T)}(1-T- q_0+T p_0).
     \label{approx}
 \end{equation}
 We emphasize that this approximate inequality has to be treated carefully and is applicable only to states for which the approximation (\ref{pqTaylor}) holds.
Employing the approximate criterion (\ref{approx}) allows us to explore the limit of highly attenuated states (\ref{rhodefinition}) with $\bar{n} \ll \eta \ll 1$. 
If we expand the exponential functions in the expressions (\ref{p0q0attenuated}) in Taylor series and insert the resulting approximate formulas for $p_0$ and $q_0$  into Eq. (\ref{approx}), we obtain 
\begin{equation}
    \eta^2 > \frac{3}{2(2-T)}\bar{n}.
    \label{modelApp}
\end{equation}
We can see that the choice of transmittance $T$ affects the ability to certify the quantum non-Gaussianity even in this limit of highly attenuated states. The formula (\ref{modelApp}) together with the plot in Fig.~3 might suggest 
that choosing $T$ close to $0$ is optimal for states (\ref{rhodefinition}). However, the statistical uncertainty imposed by finite number of measurements makes the limit $T\rightarrow 0$ impractical,
 as discussed in the following section.

\section{Comparison of single-APD and double-APD detection schemes}
The quantum non-Gaussianity criteria implied by the threshold (\ref{qpsinglemodeoptimal}) can be tested experimentally by the two detection schemes shown in Fig.~\ref{fig0}, that differ by the number of employed APDs. 
In this section, we compare the performance of those two schemes in terms of the number of measurement runs $N$ that are required to reliably certify the quantum non-Gaussianity of the probed state. 
Let us consider experimental estimation of a specific witness $W=q_0-\lambda p_0$, with threshold value $W_G$ given by Eq. (\ref{WGthreshold}). 
In the setup with a single APD, $p_0$ and $q_0$ have to be measured sequentially. Assuming that $N/2$ independent experimental runs  are spent to estimate each of the two vacuum probabilities $q_0$ and $p_0$, we get
\begin{equation}
\langle(\Delta W)^2\rangle_S=\frac{2}{N}\left[q_0(1-q_0)+\lambda^2 p_0(1-p_0)\right].
\label{deltaWsingle}
\end{equation}
Here we took into account that the measured numbers of clicks and no-clicks of the detector obey binomial distribution.
Let us now turn our attention to the double-APD scheme. Here the probabilities $p_0$ and $q_0$ are estimated simultaneously, which induces nonvanishing covariance between the two estimated statistical probabilities. 
Consider the following three mutually exclusive events: (i)  none of the two detectors clicks,  (ii) only the detector APD$_B$ clicks,  and (iii) the detector APD$_A$ clicks, irrespective of the response of APD$_B$. 
The probabilities of these three events read $p_0$, $p_{B}=q_0-p_0$, and $1-q_0$, respectively \cite{Lachman2013}, and the number of observations of these events in $N$ experimental runs obeys multinomial distribution. 
Therefore, we have
\begin{equation}
\langle \Delta p_0 \Delta  p_{B}\rangle_D=-\frac{1}{N}p_0p_{B}=-\frac{1}{N}p_0(q_0-p_0).
\end{equation}
It follows that $\langle \Delta p_0 \Delta q_0\rangle_D=p_0(1-q_0)/N$. Taking this into account, we can express the variance of the quantum non-Gaussianity witness as
\begin{equation}
\langle(\Delta W)^2\rangle_D=\frac{1}{N}\left[q_0(1-q_0)+\lambda^2 p_0(1-p_0)-2\lambda p_0(1-q_0)\right].
\end{equation}

The number of measurements $N$ required for reliable certification of quantum non-Gaussianity can be estimated by requiring that the variance of $W$ is equal to the square of the distance of the true value of $W$ from the Gaussian boundary $W_G$,
\begin{equation}
\tcr{    N_{S}=\min_\lambda \frac{\langle (\Delta W)^2)\rangle_S}{(W-W_G)^2}, \qquad
    N_{D}=\min_\lambda \frac{\langle (\Delta W)^2)\rangle_D}{(W-W_G)^2}. }
\label{NSDdefinition}
\end{equation}
\tcr{The choice of the witness, i.e. the parameter $\lambda$, is optimized to minimize the required number of measurements. 
 Note that} the threshold value $W_G$ is for each $\lambda$ (or equivalently $V$) given by Eq. (\ref{WGthreshold}), and $W=q_0-\lambda p_0$ is evaluated for the true value of the probabilities $p_0$, $q_0$. 
The advantage of the double-APD scheme can be quantified by the ratio of the required numbers of measurements,
\begin{equation}
\tcr{R_{DS}=\frac{N_D}{ N_S}.}
\label{RDSdefinition}
\end{equation}
An analytical  lower bound on $R_{DS}$ can be obtained \tcr{by assuming the same value of $\lambda$ for both detection schemes in Eq. (\ref{NSDdefinition}), which results in a simple dependence on $\lambda$,}
\begin{equation}
\fl \qquad \qquad
\tcr{R_{DS,\mathrm{min}}=\min_\lambda \frac{\langle (\Delta W)^2)\rangle_D}{\langle (\Delta W)^2)\rangle_S}=\min_\lambda \frac{1}{2}\left[1 -\frac{2\lambda p_0(1-q_0)}{q_0(1-q_0)+\lambda^2p_0(1-p_0)}\right].}
\label{RDSnewdefinition}
\end{equation}
\tcr{ Minimization of the right-hand-side of Eq. (\ref{RDSnewdefinition}) over $\lambda$ yields}
\begin{equation}
 R_{DS,\mathrm{min}}=\frac{1}{2}\left(1- \sqrt{\frac{p_0(1-q_0)}{q_0(1-p_0)}}\right)
 \label{RDSmin}
\end{equation}
\tcr{and this minimum} is attained at 
\begin{equation}
\lambda_0^2=\frac{q_0(1-q_0)}{p_0(1-p_0)}.
\label{lambda0}
\end{equation} 
It holds that $N_S \leq N_D/R_{DS,\mathrm{min}}$. This upper bound on $N_S$ is generally not tight because the optimal witnesses 
can differ for the single-APD and double-APD schemes, which means that a lower number of measurements  $N_S$ may be sufficient. Therefore, $ R_{DS,\mathrm{min}}$ represents a conservative 
estimate of the difference between the performances of the single-APD and double-APD schemes.
Two factors contribute to $R_{DS}$. First, the factor of $2$ stems from the fact that in the double-APD scheme the two vacuum probabilities $p_0$ and $q_0$ are measured simultaneously, 
while sequential measurement is required in the scheme with a single APD. Second, the positive statistical correlation between $q_0$ and $p_0$ reduces the statistical uncertainty of the witness estimation in the double-APD scheme, 
compared to the single-APD scheme. \tcr{If we choose $\lambda$ that is optimal for the single-APD scheme and use it also for the double-APD scheme, then we find that $N_S \geq 2N_D$, which implies an upper bound on  $R_{DS}$, $R_{DS}\leq \frac{1}{2}$. }

\begin{figure}[t]
\centerline{ \includegraphics[width=0.6\linewidth]{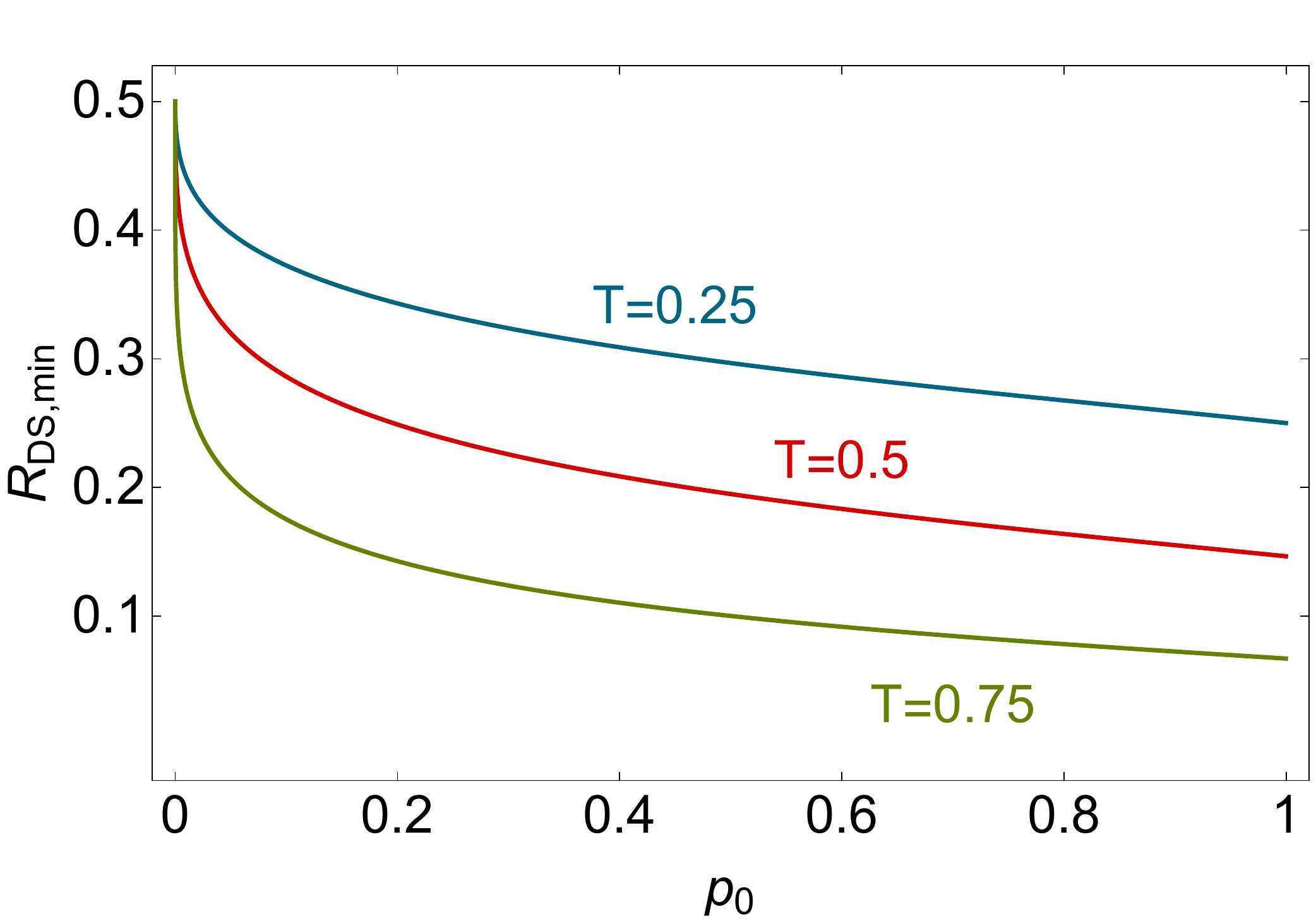}}
\caption{Dependence of $R_{DS,\mathrm{min}}$ on \tcr{the vacuum probability $p_0$ is plotted for states whose probabilities  $p_0$ and $q_0$ are equal to the boundary probability pairs} $p_0(V,T)$ and $q_0(V,T)$ achievable 
with Gaussian states. The three plotted curves correspond to three different transmittances $T=0.25$ (blue line), $T=0.5$ (red line), and $T=0.75$ (green line). }
\label{fig_RDS}
\end{figure}

We note that if one possesses \tcr{some prior estimates} of $p_0$ and $q_0$ then one may try to optimize the numbers of measurements in the single-APD 
scheme and use $K$ measurements for estimation of $q_0$ and $N-K$ measurements for estimation of $p_0$. \tcr{Such prior estimates of $p_0$ and $q_0$ can be obtained either 
from preliminary measurements on $M \ll N$ samples or from a theoretical model of the studied light source.} With this approach, formula (\ref{deltaWsingle}) changes to
\begin{equation}
\langle(\Delta W)^2\rangle_S=\frac{1}{K}q_0(1-q_0)+\frac{\lambda^2 }{N-K}p_0(1-p_0).
\label{deltaWsinglesplit}
\end{equation}
The optimal choice of $K$ that minimizes the variance is 
\[
K_{\mathrm{opt}}=\frac{\lambda_0}{\lambda+\lambda_0}N,
\]
where $\lambda_0$ is defined in Eq. (\ref{lambda0}). On inserting $K_{\mathrm{opt}}$ back into Eq. (\ref{deltaWsinglesplit}) we obtain
\begin{equation}
\langle(\Delta W)^2\rangle_S=\frac{1}{N}\left[\sqrt{q_0(1-q_0)}+\lambda \sqrt{p_0(1-p_0)}\right]^2.
\end{equation}
Interestingly, a lower bound on $R_{DS}$ derived using this modified expression for $\langle(\Delta W)^2\rangle_S$ coincides with the original bound (\ref{RDSmin}). 
This can be explained by observing that for $\lambda=\lambda_0$ we get $K_{\mathrm{opt}}=N/2$.

\begin{figure}[t]
\centerline{ \includegraphics[width=0.98\linewidth]{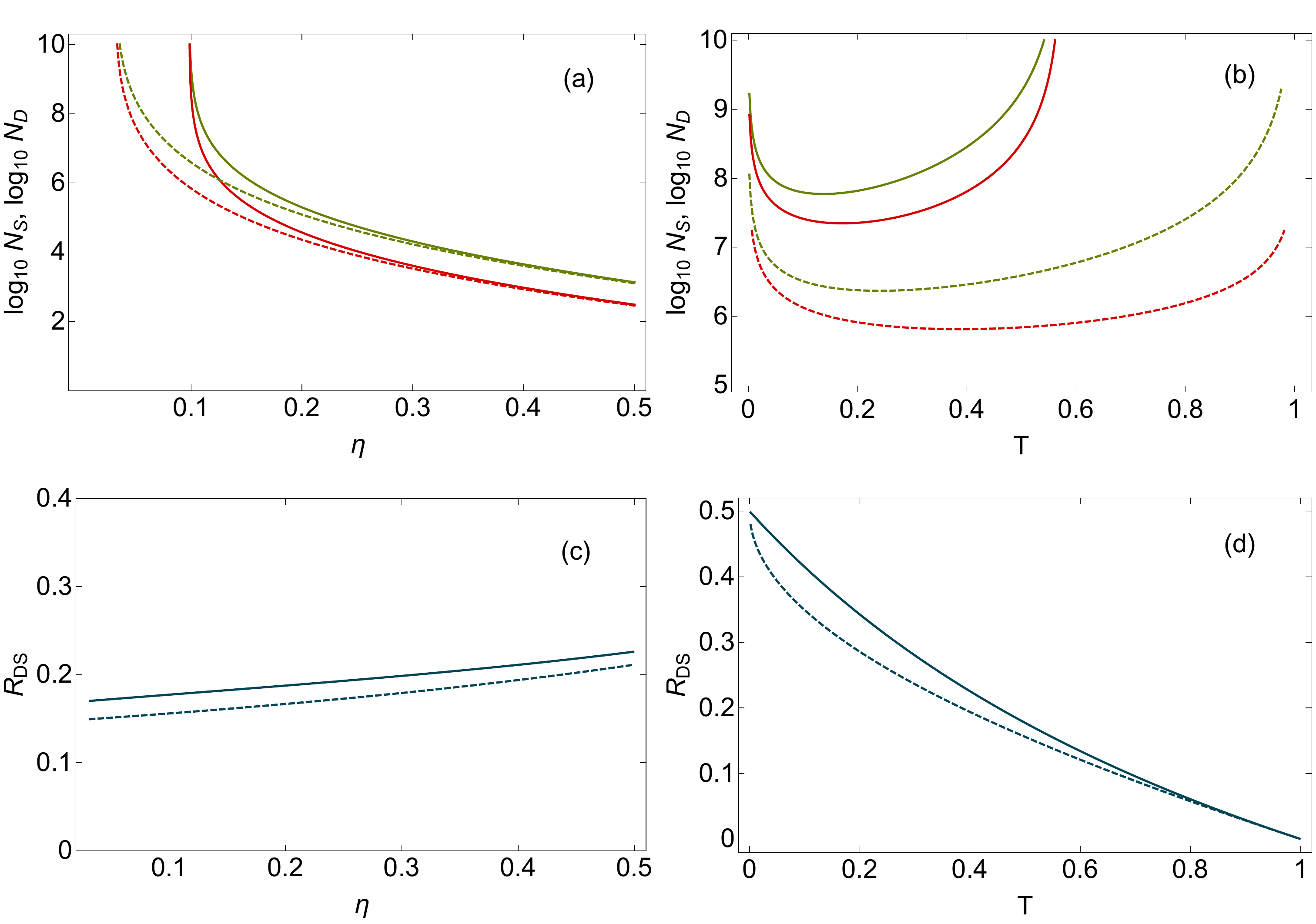}}
\caption{The number of experimental runs required for reliable certification of quantum non-Gaussianity of attenuated single-photon states with background Poissonian noise is plotted against the detection 
efficiency $\eta$ for fixed $T=0.5$ (a), and against the transmittance $T$ for fixed $\eta=0.1$ (b).  In both figures,  the green lines stand for the single-APD scheme and the red ones for the double-APD scheme. 
The results are displayed for two different levels of background noise: $\bar{n}=10^{-2}$ (solid lines) or $\bar{n}=10^{-3}$ (dashed lines). \tcr{Panels (c) and (d) show the corresponding ratios of numbers of experimental runs 
$R_{DS}=N_D/N_S$ (solid lines) as well as the lower bound $R_{SD,\mathrm{min}}$ (dashed lines). 
The curves plotted in panels (c,d) correspond to $\bar{n}=10^{-3}$ and the results for $\bar{n}=10^{-2}$ are practically identical.}}
\label{FigErrorBars}
\end{figure}

The ratio $R_{DS}$ can be in principle arbitrarily small, so the advantage of the double-APD scheme can be arbitrarily large. However, this picture changes 
if we consider only the class of states whose quantum non-Gaussianity can be certified from the measurement of $p_0$ and $q_0$, i.e. the orange areas in Fig.~2.
For a given chosen transmittance $T$ and fixed $p_0$,  $R_{DS,\mathrm{min}}$ is an increasing function of $q_0$. Therefore, for the states inside the orange area in Fig.~2, the parameter $R_{DS,\mathrm{min}}$ 
will be minimal at the boundary of this region, i.e. for the vacuum probability values  specified by Eq. (\ref{qpsinglemodeoptimal}).
In Fig.~\ref{fig_RDS} we plot the dependence of $R_{DS,\mathrm{min}}$ on \tcr{$p_0$ for these boundary states.} 
We \tcr{observe} that for \tcr{a fixed $T$  the lower bound} $R_{DS,\mathrm{min}}$ is a decreasing function of \tcr{$p_0(V,T)$} and approaches its minimum \tcr{when $p_0(V,T)\rightarrow 1$, which is equivalent to $V\rightarrow 1$. }
We can use the Taylor series expansion (\ref{pqTaylor}) to obtain a simple analytical expression for $R_{DS,\mathrm{min}}$ in the limit $V\rightarrow 1$,
\begin{equation}
R_{DS,\mathrm{min}}=\frac{1-\sqrt{T}}{2}.
\end{equation}
Let us now illustrate typical numbers of measurement runs required for certification of quantum non-Gaussianity of the attenuated single-photon states with background noise (\ref{rhodefinition}). 
In  Fig.~\ref{FigErrorBars}(a) we plot the dependence of $N_S$ and $N_D$ on the single-photon fraction $\eta$ for $T=0.5$ and two different levels of background noise $\bar{n}$. 
The plotted $N_S$ and $N_D$ are the minimal achievable values obtained by optimization over all possible witnesses.
 Fig.~\ref{FigErrorBars}(b) depicts the dependence of $N_S$ and $N_D$ on $T$ for fixed  $\eta=0.1$. 
The graphs confirm that reliable detection of the quantum non-Gaussianity with a single APD is more demanding on the number of experimental runs than the scheme with two APDs. 
The graph in Fig.~\ref{FigErrorBars}(a) illustrates that the number of required measurements grows to infinity when the true value $W$ approaches the boundary value $W_G$.
The graph ~\ref{FigErrorBars}(b) indicates that we can minimize the number of required measurements by the optimal choice of $T$. 
Furthermore, the figure also shows that the limits  $T \ll 1$ and $1-T \ll 1$ are not suitable for certification of quantum non-Gaussianity due to increasing statistical uncertainty. 
\tcr{Finally, the plots in Fig.~\ref{FigErrorBars}(c,d) show that for the considered imperfect single-photon states with background Poissonian noise the
lower bound $R_{DS,\mathrm{min}}$ provides a very good estimate of the true ratio of the required numbers of experimental runs $R_{DS}$.}

\section{Conclusions}

In summary, we have derived analytical criteria for quantum non-Gaussianity of optical quantum states  based on measurements of vacuum probabilities of the input state and state attenuated with arbitrary transmittance $T$. 
These criteria generalize the criteria previously obtained for $T=1/2$ in Ref. \cite{Lachman2013} and we have shown that our generalized criteria can be more advantageous than the criterion for $T=1/2$ 
and can detect larger class of quantum non-Gaussian states. 
\tcr{The required vacuum probabilities can be measured with setups that contain} either a single APD or two APDs. However, the more economic setup with a single APD requires larger detection time. As an interesting spin-off, 
we have also obtained a quantum non-Gaussianity criterion based on the probability of vacuum and the mean photon number. The intuition behind the investigated criteria  is that if for a given $p_0<1$  the vacuum probability 
after attenuation is large enough (or the mean photon number is low enough) then the state contains large enough fraction of the single-photon state (\tcr{or other low Fock states}) to be quantum non-Gaussian. 
We have illustrated application of the derived criteria on attenuated single-photon states with background noise which represents a realistic model of experimentally generated single-photon states. 
We have proven that our quantum non-Gaussianity criteria are applicable to general multimode states. 
This is very important for experiments with broadband single-photon detectors and single-photon sources not yet operating in the single mode regime, where single-mode detection cannot be \tcr{easily} guaranteed.

\ack
J. F. was supported by the Czech Science Foundation (GC19-19722J). L.L. was supported by the project 21-13265X of the Czech Science Foundation. 
R.F. acknowledges the MEYS of the Czech Republic and the funding from European Union’s Horizon2020 (2014-2020) research and innovation framework programme under grant agreement No 731473 (ShoQC). 

\appendix
\section{Squeezed vacuum states are not optimal}

As pointed out in Section 2, when maximizing $q_0(T)$ for a given fixed $p_0$ over the set of Gaussian states one should also consider states with zero displacement, $d_x=d_y=0$, 
i.e. the squeezed vacuum states with 
\begin{equation}
\fl
p_{0,SV}(W)=\frac{2\sqrt{W}}{W+1}, \qquad q_{0,SV}(W,T)=\frac{2\sqrt{W}}{\sqrt{(TW+2-T)(T+2W-TW)}},
\end{equation}
where $W$ is the normalized quadrature variance of the state. One can intuitively expect that this state with super-Poissonian photon number distribution cannot be optimal in the given context. Here we provide 
an explicit proof by considering an infinitezimal coherent displacement of the state combined with simultaneous infinitesimal change of $W$. We show that this change increases $q_0$ while keeping $p_0$ unchanged 
(in the first order in the infinitesimal parameter $\epsilon$). 
More specifically, consider the squeezed coherent state with infinitesimally small displacement $d_x^2=\epsilon$ and $d_y^2=0$,
\begin{equation}
\fl
p_{0}(V,\epsilon)=\frac{2\sqrt{V}}{V+1}\exp\left(-\frac{\epsilon}{V+1}\right), \quad 
q_0(V,T,\epsilon)=\frac{2\sqrt{V} \displaystyle \exp\left(-\frac{T\epsilon}{TV+2-T}\right)}{\sqrt{(TV+2-T)(T+2V-TV)}}.
\end{equation}
Let us now assume that $W=V+K\epsilon$ and seek $K$ such that 
\begin{equation}
p_{0,SV}(V+K\epsilon)=p_0(V,\epsilon)
\label{epsilonequation}
\end{equation}
holds to the first order in $\epsilon$. By expanding both sides of Eq. (\ref{epsilonequation}) as power series of $\epsilon$ and comparing terms linear in $\epsilon$ we get
\begin{equation}
K=-\frac{2V}{1-V}.
\end{equation}
Let us now compare the vacuum probabilities after passing through a lossy channel with transmittance $T$. We have
\begin{equation}
\fl
q_{0,SV}(V+\tcr{K}\epsilon,T)-q_0(V,T,\epsilon)\approx  -\frac{4(1-T)T\sqrt{V} \epsilon}{[(2-T+TV)(2V+T-TV)]^{3/2}} <0.
\end{equation}
This proves that for any given fixed $p_{0}$ and $T$ a suitably infinitesimally displaced squeezed coherent state exhibits larger $q_0(T)$ than the squeezed vacuum state.
Therefore, the squeezed vacuum states are not optimal and it is sufficient to consider only squeezed coherent states with nonzero $d_x$ as is done in the final optimization step in Section 2.

\section{Multimode optimality proof}

In many experiments with broadband single-photon emitters and avalanche photodiodes or similar single-photon detectors it is usually very hard or impossible to satisfy the condition that only a single mode is detected. 
Although the spatial and polarization modes can be filtered, there remain temporal (or spectral) modes.
In order to apply the quantum non-Gaussianity criteria to such experiments, we have to prove that they hold also for multimode states.
Our proof is based on similar earlier proofs outlined in Refs. \cite{Jezek2011,Lachman2013} but we provide additional technical details and use approaches that allow us to prove the criteria for arbitrary $T$.

For general multimode states  the witness $W=q_0(T)-\lambda p_0$ can be expressed as $W=\mathrm{Tr}[\hat{\rho}\hat{W}_{T,N} ]$, where $\hat{W}_{T,N}$ is an operator diagonal in multimode Fock state basis.  
Therefore, when seeking the maximum of $W$ over Gaussian states and their mixtures, it suffices  to perform the optimization over pure $N$-mode Gaussian states. 
In other words, pure states are optimal also in the general multimode case. The next important observation is that product pure Gaussian states are optimal. 
Any pure $N$-mode Gaussian state can be transformed to a product pure $N$-mode Gaussian state by a passive unitary Gaussian operation that does not  modify the total photon number distribution 
and therefore does not change $p_0$ and $q_0(T)$. Physically, this transformation is represented by a suitably designed $N$-port passive linear optical interferometer \cite{Braunstein2005}. 
For an arbitrary product $N$-mode state, the vacuum probabilities can be expressed as 
\begin{equation}
p_0=\prod_{k=1}^N p_{0,k},\qquad q_0(T)=\prod_{k=1}^N q_{0,k}(T),
\end{equation}
where the index $k$ labels the modes and $p_{0,k}$ and $q_{0,k}(T)$ denote the single-mode vacuum probabilities for $k$-th mode.
Since our goal is to maximize $q_0(T)$ then the state in each mode $k$ should be chosen such that for a given $p_{0,k}$ it maximizes $q_{0,k}(T)$. It thus suffices to consider product 
Gaussian state where each single-mode state is the optimal state specified by Eq. (\ref{qpsinglemodeoptimal}).

 Let us first consider the two-mode case. Labeling the two modes with letters $A$ and $B$, we can write 
\begin{equation}
W= q_{0}(V_A,T)q_{0}(V_B,T)-\lambda p_{0}(V_A,T) p_{0}(V_B,T),
\end{equation}
where $V_A$ and $V_B$ denote the normalized quadrature variances of modes $A$ and $B$, respectively. The extremal equations read
\begin{equation}
\frac{\partial W}{\partial V_j}=0, \qquad j=A,B.
\end{equation}
Following the same procedure as in Section 2, we can eliminate the Lagrange multiplier $\lambda$, which yields
\[
\fl
p_{0}(V_A,T)q_{0}(V_B,T) \frac{\partial p_{0}(V_B,T)}{\partial V_B} \frac{\partial q_{0}(V_A,T)}{\partial V_A}-
p_{0}(V_B,T)q_{0}(V_A,T) \frac{\partial p_{0}(V_A,T)}{\partial V_A} \frac{\partial q_{0}(V_B,T)}{\partial V_B}=0.
\]
After some algebra, we obtain the following equivalent equation,
\begin{equation}
T(1-T)(V_A-V_B)P(V_A)P(V_B)=0,
\label{twomodeoptimality}
 \end{equation}
where 
\begin{eqnarray}
P(V)&=&-(2-T)^2 V^4-2(1-T)T V^3+2(2-T)(T+1) V^2 \nonumber \\
 & & +2(4-(3-T)T)V +(2-T) T.
\end{eqnarray}
We show that for any $T\in(0,1)$ the polynomial $P(V)$ does not have any real root in the interval $(0,1]$. For $V\in (0,1]$ we have
\begin{eqnarray}
 P(V)&=&V^2\left[-(2-T)^2 V^2-2(1-T)T V+2(2-T)(T+1)\right. \nonumber \\
 & & \left. +\frac{2(4-(3-T)T)}{V}+\frac{(2-T) T}{V^2}\right] \nonumber \\
& \geq& V^2 \left[-(2-T)^2 -2(1-T)T +2(2-T)(T+1) \right. \nonumber \\
& &\left. +2(4-(3-T)T)+(2-T) T\right]  = 8 V^2>0.
\end{eqnarray}
Here we took into account that $0<T<1$. It follows that the optimality condition (\ref{twomodeoptimality}) can be satisfied only if  $V_A=V_B$. In this case we have two identical pure squeezed coherent states in modes A and B. 
We can formally transform to superposition basis by combining the modes A and B on a balanced beam splitter,  
which does not change the probabilities $p_0$ and $q_{0}(T)$. After this formal transformation, we again have a product two-mode pure Gaussian state with single-mode squeezing 
$V=V_A=V_B$ identical in both modes. However, due to destructive and constructive interference, one mode will have zero coherent displacement and the other mode will have displacement amplified by factor of $\sqrt{2}$. 
 Clearly, these latter single-mode Gaussian states are not the optimal states (\ref{qpsinglemodeoptimal}). Therefore, the case $V_A=V_B$ cannot lead to maximization of $q_0(T)$ for a fixed $p_{0}$.

The above analysis shows that the extremality condition (\ref{twomodeoptimality}) yields a single candidate point $V_A=V_B$ that however does not correspond to the sought maximum. 
Therefore, the maximum of $W$ is instead reached at the boundary of the allowed range of parameters $V_A$ and $V_B$, where either $V_A=1$ or $V_B=1$. Thus, in the two-mode case 
the optimal state is actually the optimal single-mode state in mode A and the vacuum state in mode B. Hence the quantum non-Gaussianity criteria for 
the two-mode case are exactly the same as for the single-mode case.

Finally, we extend our proof to an arbitrary number of modes by mathematical induction. Let us assume that for an $N$-mode case it holds that the quantum non-Gaussianity 
criterion is the same as for the single-mode case and the optimal pure Gaussian state maximizing the witness $W$ is a product of $N-1$ vacuum states and the optimal single-mode squeezed coherent state (\ref{qpsinglemodeoptimal}). 
For the $N+1$-mode case we can then write 
\begin{equation}
p_{0}^{(N+1)}=p_{0}^{(N)}p_{0}^{(1)}, \qquad  q_{0}^{(N+1)}(T)=q_{0}^{(N)}(T)q_{0}^{(1)}(T),
\end{equation}
 where $p_{0}^{(N)}$ and $q_{0}^{(N)}(T)$ are the optimal probabilities for the $N$-mode case. Since for the \tcr{$N$-mode} case the optimal probabilities are those of the single-mode case, 
 the $N+1$-mode optimization problem reduces to the two-mode case. This has been solved in the previous step where it was shown that the optimal choice is vacuum in one mode 
 and the optimal single-mode squeezed coherent state in the other mode. This proves the assumption also for the case of $N+1$ modes. 
 By induction, the result holds for an arbitrary number of modes and the quantum non-Gaussianity criteria derived in Sections 2 and 3 of the present paper are applicable to general multimode states.

\end{document}